\documentstyle[aps,multicol,prb,epsf]{revtex}
\begin{document}
\title{T--Matrix Formulation of
Impurity Scattering in Correlated Systems}
\author{ W.\ Ziegler, D.\ Poilblanc$^{\bf *}$,
          R.\ Preuss ,W.\ Hanke
 and D.\ J.\ Scalapino $^{\bf \dagger}$\\
\medskip
\em     Institut f\"ur Theoretische Physik, Universit\"at W\"urzburg \\
           97074 W\"urzburg, FRG. \\
      $^{\bf *}$Laboratoire de Physique Quantique,
      Universit\'e Paul Sabatier,\\
         31062 Toulouse, France.\\
$^{\bf \dagger}$ Department of Physics, University of California at
Santa Barbara,\\
Santa Barbara, California 93106.\em }
\maketitle
\begin{abstract}
Using a generalized T-matrix description which, in principle, exactly includes
Coulomb correlations and potential scattering events, resonant and bound
impurity states are discussed.
Like in the non-interacting case, the effects of the scattering potential
can be divided into different partial wave channels, exploiting the symmetry
of the underlying lattice. Due to Coulomb correlations bare local
(i.e.\ s-wave) potentials
become dynamic and extended, being responsible also for
p-, d-wave etc. scattering effects. Numerically exact results for both the
two-dimensional t--J and Hubbard models are used to construct a simple (static)
approximation to the effective impurity potential which is shown to reproduce
the
exact resonant scattering and bound states in the relevant symmetry channels.
\end{abstract}
\begin{multicols}{2}
%
\indent
Experimental data on high--$T_c$ materials including substitutional defects
serve for probing their unusual normal state properties as well as the nature
of the superconducting phase.
Introducing, for example, Zn in the copper oxide planes
influences transport and magnetic properties
and leads to a drastic
reduction of the transition temperature $T_c$
\cite{Xiao90,Chien91,Zhao93}.
The physical conclusions drawn from the
results are still  controversial. For example,
this is the case for the local magnetic
moment induced by a non--magnetic impurity such as Zn in hole doped
superconducting samples \cite{Xiao90,Fink90,Alloul91,Cieplak92,Ishida93}
which could lead to a magnetic pair breaking mechanism
\cite{Walstedt93,Mahajan94}.\\
\indent
However, one \em common \em feature seems to be the  spatially
extended nature of the
magnetic effects on the $CuO$--based  high--$T_c$ compounds.
{}From residual resistivity measurements  one deduces large scattering
cross sections of several lattice constants in diameter \cite{Chien91,Zhao93}
favoring strong potential scattering  likely connected to d--wave
pair breaking \cite{Walstedt93}.
%
On the theoretical side there has been work based on s--wave
impurity scattering in
anisotropic superconductors revealing the strong influence of
non--magnetic impurities
\cite{Hirschfeld88,Millis88,Hotta92,Borkowski94,Fehrenbacher94}.
In addition, calculations were carried out for d--wave  superconductors
with  combined scattering by  impurities and Coulomb correlations
\cite{Quinlan94,Hirschfeld94}.\\
\indent
In this paper we focus on the
interplay between strong correlations and local potential scattering.
This situation can be ascribed to  Zn$^{2+}$ impurities with a filled
(3d)$^{10}$ shell in the cuprate superconductors.
We derive a  microscopic picture of the mediated scattering processes
using a generalized T-matrix description which exactly includes
 Coulomb correlations
and which exploits the symmetries of the underlying lattice.
Partial-wave phase shifts
are introduced which, like in the non-interacting case determine the additional
density of states caused by the impurity potential.
For this purpose we take advantage of numerical exact results of
translational--invariant
lattices and combine  them in analytical T--matrix expressions
for a single impurity.
By doing this we can determine the effective potential strengths acting
in a perturbed $t$--$J$ system investigated earlier \cite{Imptj}.\\
\indent
%
Consider a correlated two-dimensional system such as the one  band
Hubbard  model with Hamiltonian,
\begin{equation}
   H\,=\,
   H_0 + H_U\,=
   \,-t
   \sum\limits_{\stackrel{\langle i,j \rangle}{\sigma}}
    c_{i\sigma}^{\dagger}c_{j\sigma}^{ }
    +
    U\sum\limits_{i}n_{i\uparrow}n_{i\downarrow}
    ,
    \label{Ham}
\end{equation}
using standard notations,
and a bare impurity operator
\begin{equation}
V_{Imp,\sigma}\,=
\, \sum_{\delta\sigma}
 (t_{0\sigma}c_{0\sigma}^{\dagger}c^{}_{\delta\sigma}+ h.c.)
+\sum_\sigma V_{0\sigma}^{bare} n_{0\sigma}.
\end{equation}
Here the four bonds connected to the impurity site
at the origin are modified by $t_{0\sigma}$.
The simplest description of a vacancy is then obtained by
setting $t_{0\sigma}=t$.
For the sake of simplicity the potential is restricted here to the central
site but the following analysis is more general and valid for extended one-body
potentials.
The above scattering problem on a lattice is defined
spin--dependent and  can therefore be adapted to
different physical scenarios.
In  the case of non--interacting
systems, e.g.\ $U=0$, successive scattering is described by the
Greens function
$G_{\sigma}=G_{0} + G_{0}\,T_{\sigma}\,G_{0}$.
Here, the  T--matrix,  defined as in usual scattering theory,
$T_{\sigma}=V_{Imp,\sigma}\,( 1-G_{0}\,V_{Imp,\sigma})^{-1}$,
involves only the unperturbed propagator $G_{0}$ and the bare
potential $V_{Imp,\sigma}$.  \\
On the other hand, for correlated systems it is $ad$ $hoc$ not clear
how to define a T--matrix  leading to a similar result for the
exact Greens function $G$.
This can be achieved by investigating the representation
of $G$ extracted from perturbation theory.
For  finite temperatures, the exact  Greens Function $G$
can be written in the Matsubara formalism and in momentum space as
\begin{equation}
G_{\sigma}^{-1}({\bf k},{\bf k}^{\prime},\omega_n)
=i\omega_n -\varepsilon({\bf k})
-\Sigma_{\sigma}({\bf k},{\bf k}^{\prime},\omega_n).
\label{Gtot}
\end{equation}
The self--energy $\Sigma$ is a sum of three terms
$\Sigma_U({\bf k},{\bf k},\omega_n)$,
$\Sigma_{UV,\sigma}({\bf k},{\bf k}^{\prime},\omega_n)$
and $V_{Imp,\sigma}({\bf k},{\bf k}^{\prime})$.
In a diagrammatic analysis $\Sigma_{U}({\bf k},{\bf k},\omega_n)$
contains exclusively many--body effects,
the mixed self--energy term
$\Sigma_{UV,\sigma}({\bf k},{\bf k}^{\prime},\omega_n)$
includes all correlation diagrams with an internal line scattered at least once
by $V_{Imp,\sigma}$ \cite{Langer60},
examples of which are given in Figs.\ 1b) and 1c).
$V_{Imp,\sigma}({\bf k},{\bf k}^{\prime})$
 is then just a single scattering event
caused by the bare impurity potential.
Defining a dynamical effective potential
\begin{equation}
V_{eff,\sigma}({\bf k},{\bf k}^{\prime},\omega_n)=
\Sigma_{UV,\sigma}({\bf k},{\bf k}^{\prime},\omega_n)
+ V_{Imp,\sigma}({\bf k},{\bf k}^{\prime})
\label{Veff}
\end{equation}
one arrives at the desired formulation, i.e.
\begin{eqnarray}
\lefteqn{G_{\sigma}({\bf k},{\bf k}^{\prime},\omega_n) =
 G_U({\bf k},{\bf k},\omega_n)  } \nonumber \\
&+&
G_U({\bf k},{\bf k},\omega_n)\,T_{\sigma}({\bf k},{\bf k}^{\prime},\omega_n)\,
G_U({\bf k}^{\prime},{\bf k}^{\prime},\omega_n),
\end{eqnarray}
where
$
G_U^{-1}({\bf k},{\bf k},\omega_n)=
i\omega_n -\varepsilon({\bf k})
-\Sigma_U({\bf k},{\bf k},\omega_n)
$
and, by omitting the momentum indices,
\begin{equation}
T_{\sigma}(\omega_n)=
V_{eff,\sigma}(\omega_n)\,\left(1-G_U(\omega_n)\,
V_{eff,\sigma}(\omega_n)
\right)^{-1}.
\label{Tmatrix}
\end{equation}
\indent
In order to investigate the one--particle behavior of the system by means
of the T--matrix, one exploits the transformation of its  matrix elements
in real space according to the symmetrized  states of the underlying
point-group symmetry of the lattice \cite{Callaway}. This means
that the effects of the scattering potential,
which respects this symmetry, can be divided
into  contributions to different partial--wave channels.
Note that for the non--interacting case,
the bare potentials $V_{0\sigma}^{bare}$ and
$t_{0\sigma}$ only contribute to s--wave processes, while scattering
in the d--, p--wave etc. channels would require extended bare potentials.
As seen later, this is substantially
modified by the Coulomb interaction
leading to correlation induced longer-ranged effective potentials.
\indent
The poles of the  retarded T--matrix, which
correspond to the zero--values of its  determinant,
either determine resonant states, which
are found inside the bands of the unperturbed
($V_{Imp}=0$) system, or bound states separated from
the continuum. These statements are clear for non--interacting
systems \cite{Callaway} but  hold also for $U\neq 0$.
Note that
$V_{eff,\sigma}(\omega)$ is a real quantity
at the bound state energies  $\omega=\omega_{BS}$
due to the infinite lifetime of these states.

\indent
By using the irreducible representations
the complex determinant of the T--matrix
factorizes into the subdeterminants
of its partial--wave decompositions.
The partial--wave phase shifts $\phi_{\alpha}(\omega)$
are then introduced as
the (negative) phase of these $\alpha$--wave subdeterminants.
They determine the additional
density of states (DOS) caused by an extended impurity potential
\cite{Callaway}.
\indent
The total DOS of the system is  given by the trace over the
retarded Greens function $G^{ret}$
and can be written as the sum of  $D_U(\omega)$,
the DOS  of the correlated system without the impurity
and an additional impurity--caused contribution $\Delta D_{\sigma}(\omega)$.
Bringing in the phase shifts, this leads to
\begin{eqnarray}
\Delta D_{\sigma}(\omega)&
=&-\frac{1}{N\pi}{\rm Im} \frac{\partial}{\partial \omega}
\ln\, {\rm Det}(1-G_U^{ret}(\omega)\, V_{eff,\sigma}(\omega))
, \nonumber\\
&=&
\sum_{\alpha}  \Delta D_{\alpha,\sigma}(\omega)
=
\frac{1}{N\pi}\sum_{\nu} g(\alpha)
\frac{\partial\phi_{\alpha,\sigma}}{\partial \omega},
\label{AddDOS}
\end{eqnarray}
where $N$ counts the number of sites and $g(\alpha)$ denotes the
degeneracy of the representation $\alpha$.\\
\indent
This shows that the phase shifts and the T--matrix defined
for non-interacting systems are still helpful concepts for studying
bound states and resonant scattering in the interacting case. However,
the use of Eq.(\ref{Tmatrix}) implies the knowledge of
the {\it effective} potential which is {\it a priori}
frequency and interaction (U) dependent.
Note that $V_{eff}$ can be expressed as
$V_{eff,\sigma}(\omega)= {G_U^{ret}(\omega)}^{-1}
-{G^{ret}_{\sigma}(\omega)}^{-1}$,
which, in principle, would make it possible to extract
the dynamical potentials by numerical calculations.
Here we adopt a somewhat simpler approach
by showing that a crude static but nevertheless interaction-dependent Ansatz
can be made for $V_{eff,\sigma}$ which can
describe, within reasonable accuracy, the scattering processes.
This method enables us to understand in more physical terms the interplay
between the interaction and short-range impurity scattering.\\
\indent
One important feature which has to be included in the Ansatz
is the {\it longer range character} of $V_{eff}$
due to the background of the correlated host.
This can simply be seen from the connection between the bare and
the dynamically modified potential.
By explicitly transforming the Coulomb term to the real-space
irreducible representation, one finds
local interactions between different
wave--symmetries, i.e.\
\begin{equation}
H_U=
\sum_{\nu,\alpha_1,\alpha_2,\alpha_3,\alpha_4}\,\,
U_{\nu}^{\alpha_1 \alpha_2 \alpha_3 \alpha_4} \,
\,
{c_{\nu\,\uparrow}^{\alpha_1}}^{\!\dagger}\,
{c_{\nu\,\uparrow}^{\alpha_2}}^{ }\,
{c_{\nu\,\downarrow}^{\alpha_3}}^{\!\dagger}\,
{c_{\nu\,\downarrow}^{\alpha_4}}{ }
\label{Uirr}
\end{equation}
where $c_{\sigma\,\nu}^{\alpha (\dagger)}$ are the
symmetrized annihilation (creation) operators of the
one--particle spin ($\sigma$) states.
$\alpha $ labels in general all irreducible
representations of the site--type $\nu$ \cite{Anm1}.
Note that, although it contains $\alpha$-wave operators the
above expression for $H_U$ is invariant under all the
point group symmetries. This implies selection rules
on the matrix elements
$U_{\nu}^{\alpha_1 \alpha_2 \alpha_3 \alpha_4}$
which, for simplicity, are not specified
here. Therefore, even for a pure $s$--wave potential
$V_0^{bare}\sum_{\sigma}
c_{\sigma 0}^{s \dagger}c_{\sigma 0}^{s} $
at the
origin, incoming $d$--wave propagators are scattered at $\nu=1$
from a potential mediated by the Coulomb repulsion as shown in
\mbox{Fig.\ 1.}\\
\noindent
%
\indent
This demonstrates that, due to many--body effects,
even local ($s$--wave) potentials always become
dynamically \em extended\em, being in general responsible for
scattering contributions in all symmetry channels.
In our numerical results  the effective potential in Eq.(4)
is described by assuming the following static
approximation for the effective potential in Eq.(4);
\begin{equation}
V_{eff}^{stat}\,=
\, t^\prime\sum_{\sigma} (c_{0 \sigma }^{s\dagger}c_{1 \sigma }^s + h.c.)
+ \sum_{\nu\alpha\sigma} V_{\nu\sigma}^\alpha \,
c_{\nu\sigma}^{\alpha\dagger}c_{\nu\sigma}^{\alpha} .
\label{Stat}
\end{equation}

Generally, the $\nu$--off-diagonal
 (diagonal) part of $V_{eff}$ also takes into account
the effective change of the hopping amplitudes (potential terms).\\
\indent
Next, we proceed with our  numerical results.
At zero temperature,
the retarded Greens function $G_U$ appearing in the above
equations is calculated   by  exact diagonalization (ED)
techniques for the half--filled 2D $t$--$J$ model \cite{did1}.
In this context, we regard this model
\cite{Imptj} as the strong--coupling  limit
of our original Hamiltonian Eq.(\ref{Ham}) to perform the T--matrix
calculations (i.e. $J=4t^2/U$).
We consider the situation of reference \cite{Imptj}, where
an inert site in  a $t$--$J$ lattice produced bound states of
different wave symmetries $\alpha$ in the singlet one--particle excitation
spectrum. The above formalism
is  used to determine the effective potential strengths causing
these bound states.
The description in terms of spin--dependent (effective) potentials
takes into account that an excess--spin lifts the
spin--degeneracy of the system,
leading also to an antiferromagnetic structure in the vicinity of the
impurity \cite{Imptj,Bulut89}
In contrast,  an infinitesimal magnetic coupling
of the inert site to the lattice
restores the spin--degeneracy
\cite{ImptjII}, which would lead  again to spin--independent
effective potentials.
Keeping in mind,  that we  restrict our
calculations to  the low--lying excitations
in the above mentioned singlet channel \cite{Imptj},
we omit spin indices from now on.
For the case of interest here ($t_0=t$ or more generally for
$V_0^{bare}=0$) the problem becomes particle-hole symmetric.
Then, for simplicity, we define
$\omega=0$ as the lower edge of the quasiparticle band
of the $t$-$J$--model (lower Hubbard band).
In the {\it hole} representation, positive (negative) values of
$V_{0,1,2}^\alpha$ correspond to repulsive (attractive)
potentials.

According to the local DOS results of
Ref. \cite{Imptj}, the values of $V_0^{\alpha}$ and $V_2^{\alpha}$ are set to
large values $\sim 20t$ in order to expel particles from the
corresponding site--types $\nu=0,2$.
In our calculations $t^{\prime}$ increases  the bound-state
energy quadratically.
Taking the actual small binding energies of \cite{Imptj} into account
the small parameter $t^{\prime}$ is therefore set to zero.
In this sense the derived potential strengths
represent an upper limit.

Fig.\ 2 displays the exact
bound state energies  $\omega_{BS}$,
calculated from the local DOS for a
20--site cluster with an isolated site at the origin \cite{Imptj}.
By using the T--matrix, the energetic locations
of the bound states are reproduced for  values
of  $J=0.5$ and  $J=1.0$ fixing the effective
potentials  $V_1^\alpha$.
Fig.\ 2 also   exhibits the  dependence  of $V_1^\alpha$ on the
exchange coupling $J$ of the unperturbed system.
As one expects, the attractive  potentials increase with J.
One finds the actual value of the effective potential to
be of the order of J, which is the loss of magnetic energy
per bond of the inert site.\\

Due to the  dynamic renormalization, the potential $V_1^\alpha$
acts, as mentioned, differently in each symmetry sector being largest for
the s--wave channel.
Note that in the limit of vanishing interaction strength ($U\rightarrow 0$),
the $p$-- and $d$--wave  potentials also have to  vanish,
reflecting non--extended potential scattering.
Therefore, in an interacting system,
there should exist an optimum correlation strength
producing the largest effective potentials for bound states.\\
\indent
For $J=0.5$ Fig.\ 3a) shows the DOS of the unperturbed 26--site cluster
used to calculate  the additional impurity--induced density in Fig.\  3b).
Assuming the effective potentials to be  slowly  varying functions
of $\omega$ one finds a suppression of density
for small energies in agreement with the exact result
\cite{Imptj}  in Fig.\ 3c).\\

\indent
Now, we investigate scattering in the 2D Hubbard model.
For finite temperatures, we exploit  results of
Quantum Monte Carlo (QMC)/Maximum Entropy
calculations \cite{lph94} for the undoped system
by inserting the
unperturbed Greens functions into the T--matrix equations.\\
\indent
At half--filling and for $U=8t $, $\beta t=10$, we use the effective potentials
derived  from the $t$-$J$--model for $J=4t^2/U=0.5$
and for the particle-hole symmetric case $t_0=t$.
In this case, $V_0^{bare}$ is irrelevant and can be set to zero, while
the effective potentials change sign at the chemical potential $\mu$.
Both pure systems are  characterized by
long--range antiferromagnetic correlations
being disrupted by the impurity which  should lead to a
similar behavior regarding the formation of bound states.
Actually, the states appearing below the $t$-$J$ quasiparticle band
are  now located just above the
small dispersive quasiparticle band of width $\sim 2J$
riding on top of a several $t$ wide incoherent background \cite{lph94}.
This  can be seen
in Fig.\ 4 displaying the unperturbed DOS $D_U$ obtained from
QMC calculations and the (non--normalized) additional density $\Delta
D_{\alpha}$.
Differing  from the  $t$-$J$--case the s-wave contributions are shifted
towards the quasiparticle band.
The appearance of these states results in an
overall effect of  narrowing the gap of the unperturbed
Hubbard model by almost $20$\%.\\
\indent
In summary, we have considered a formally exact T--matrix
method adapted to correlated
systems which revealed the many--body interactions as the origin
of dynamically extended potentials.
Using numerical data for the unperturbed propagators,
we reproduced  the static effective potentials causing bound states
in a 2D $t$-$J$--lattice with an inert site.
Applying the formalism to the insulating 2D Hubbard model,
we have  shown the correspondence
of the two models regarding the appearance of spectral weight
in the correlation gap.
Our  result is relevant  for T--matrix approximations
in  the  dilute impurity limit, i.e.\ where the single impurity
scattering has to be treated exactly. There, higher symmetry channels
than the usually considered $s$--wave contributions have to be
taken into account for interacting systems.
Our static, effective potentials in Eq.(9) can, for example, directly
be used as an input to a cluster diagonalization approach, where the effect
of potential impurity scattering in a 2 D superconductor described by a BCS
mean--field theory \cite{Xiang94} is studied.
In ref.\ \cite{Xiang94} the impurity potential was  parameterized.
Nevertheless, already this work demonstrated that the range of the impurity
potential is of quantitative importance in the case of strong potential
scatterers.\\

\indent
D.\ P.\ and W.\ H.\ acknowledge hospitality and support from the
Physics Department Santa Barbara.
The QMC calculations were performed at HLRZ J\"ulich and at LRZ M\"unchen.
D.\ P.\ also thanks IDRIS (Orsay) for providing
CPU time for the ED calculations.\\\
%

\end{multicols}

\begin{thebibliography}{99}
\bibitem{Anm1} For the $C_{4V}$--lattice of the
	       Hubbard model there
	       exist in general five different representations.
	       At the considered
	       impurity site--types $\nu=0,1,2$,
	       only
	       $s$--, $p$-- and
	       $d$--wave representations contribute.
	       The  $p$--wave is degenerated
	       ($p_x =  p_y$).
	       Different site--types $\nu$ are distinguished
	       by their distance from the origin.
\end{thebibliography}
\end{document}